\documentclass[aps,pre,twocolumn,showpacs]{revtex4}
\usepackage{epsfig}
\usepackage{times}
\begin{document}

\title{Phase diagrams for an evolutionary Prisoner's Dilemma game \\
on two-dimensional lattices}

\author{Gy\"orgy Szab\'o$^1$, Jeromos Vukov$^2$, and Attila Szolnoki$^1$}
\affiliation
{$^1$Research Institute for Technical Physics and Materials Science
P.O. Box 49, H-1525 Budapest, Hungary  \\
$^2$Department of Biological Physics, E\"otv\"os
University, H-1117 Budapest, P\'azm\'any P. stny. 1/A., Hungary}

\begin{abstract}
The effects of payoffs and noise on the maintenance of cooperative behavior are studied in an evolutionary Prisoner's Dilemma game with players located on the sites of different two-dimensional lattices. This system exhibits a phase transition from a mixed state of cooperators and defectors to a homogeneous one where only the defectors remain alive. Using systematic Monte Carlo simulations and different levels of the generalized mean-field approximations we have determined the phase boundaries (critical points) separating the two phases on the plane of the temperature (noise) and temptation to choose defection.  In the zero temperature limit this analysis suggests that the cooperation can be sustained only for those connectivity structures where three-site clique percolation occurs.   
\end{abstract}

\pacs{89.65.-s, 05.50.+q, 02.50.+Le, 87.23.Ge}

\maketitle

In the original (two-player and one-shot) Prisoner's Dilemma (PD) game \cite{weibull_95,gintis_00} the players should simultaneously choose between two options, called defection and cooperation. The selfish players wish to maximize their own income in the knowledge of payoffs dependent on their choices. The curiosity of PD game is hidden in the fact that the choice of defection yields higher income independently of the partner's choice. However, if both players choose defection then their individual income is lower than those obtained for mutual cooperation when the maximum total payoff is shared equally. The rational (intelligent) players cannot resolve this dilemma and both of them choose defection (this is the so-called Nash-equilibrium in the PD game). At the same time we find many examples in the nature where the mutual cooperation (altruism, ethical norms, etc.) emerges spontaneously among the selfish individuals \cite{nowak_sa95}. In the last decades several mechanisms (e.g., kin selection \cite{hamilton_jtb64b}, application of retaliating strategies \cite{axelrod_84}, and voluntary participation \cite{hauert_s02}) are reported which enforce the appearance of cooperation in the societies. 

The spatial versions \cite{nowak_ijbc93,nowak_ijbc94} of the evolutionary PD games can explain the maintenance of cooperation for the iterated games with a limited range of interaction if the players follow one of the two simplest strategies. For the two simplest strategies, denoted shortly as $D$ and $C$, the player choose always defection and cooperation, respectively. In the evolutionary games the players wish to maximize their total payoff, coming from PD games with the neighbors, by adopting one of the more successful strategies available in their neighborhood. This type of dynamics describes the behavior of the ecological systems controlled by the Darwinian selection \cite{maynard_82,hofbauer_98}.

Following the pioneering work of Nowak {\it et al.} \cite{nowak_ijbc93,nowak_ijbc94} the two-strategy spatial evolutionary PD games have already been studied by several authors using different evolutionary rules on a large class of backgrounds including social networks \cite{abramson_pre01,ebel_pre02,kim_pre02,masuda_pla03} (for a survey of lattice models see the papers \cite{lindgren_pd94,nakamaru_jtb97,hauert_ajp05} and further references therein). In the present paper our attention is focused on the effect of noise built into the dynamical rule. It is turned out that the effect of noise on the stationary concentration of cooperators depends strongly on the topological features of the neighborhood and the measure of cooperation can be enhanced by increasing the noise in some cases.

For this purpose we consider an evolutionary PD game with players located on the sites $x$ of a two-dimensional lattice. The players follow one of the above mentioned two strategies whose distribution is described by a two-state Potts model, i.e., ${\bf s}_x=C$ or  $D$, where for later convenience the states are denoted by the two-dimensional unit vectors,
\begin{equation}
\label{eq:cd}
D= \left( \matrix{1 \cr 0 \cr }\right) \;\; \mbox{and}\;\;
C= \left( \matrix{0 \cr 1 \cr }\right) \;.
\end{equation}
In this notation the total income of player $x$ can be expressed as
\begin{equation}
\label{eq:tpo}
U_x=\sum_{\delta } {\bf s}^{+}_x {\bf A} \cdot {\bf s}_{x + \delta },
\end{equation}
where ${\bf s}^{+}_x$ denotes the transpose of the state vector ${\bf s}_x$, the summation runs over those four neighbors who the player $x$ plays PD game with. Following Nowak {\it et al.} \cite{nowak_ijbc93} the rescaled payoff matrix is given as 
\begin{equation}
%\label{eq:pom}
{\bf A}=\left( \matrix{0 & b \cr
                       c & 1 \cr} \right)\;, \;\; 1 < b < 2-c \;, \;\; c < 0 \;.
\end{equation}
The evolutionary process is governed by random sequential strategy adoptions, that is, the randomly chosen player $x$ adopts one of the (randomly chosen) neighboring strategy (at site $y$) with a probability depending on the payoff difference
\begin{equation}
\label{eq:update}
W[{\bf s}_x \leftarrow {\bf s}_y] = {1 \over 1 +
 \exp {[(U_x-U_y)/K]} } \;,
\end{equation}
where $K$ is the measure of stochastic uncertainties (noise) allowing the irrational choices \cite{blume_geb93,szabo_pre02d}. 

\begin{figure}[ht]
\centerline{\epsfig{file=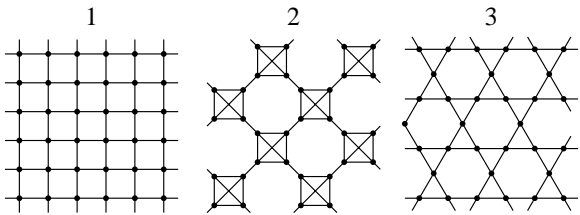,width=8cm}}
\caption{\label{fig:stages3}Three two-dimensional lattice structures on which an evolutionary Prisoner's Dilemma game is studied.}
\end{figure}

Our analysis will be restricted to two-dimensional lattices where the topologically equivalent sites have four neighbors ($z=4$) as indicated by the edges in Fig.~\ref{fig:stages3}. By this way we can avoid the undesired effects due to the variation of the number of co-players \cite{nowak_ijbc93,nowak_ijbc94,hauert_prsb01,schweitzer_acs03}. 

The investigated connectivity structures are the square (1) and Kagome (3) lattices, and a square lattice of four-site cliques (2). The latter structure consists of four-site cliques (within a clique the nodes are linked to each other) whose sites are connected only to one external site belonging to the nearest clique. These structures can be distinguished topologically by considering the connectivity between the triangles (three-site cliques). In fact, the square lattice is free of triangles, that is, its clustering coefficient ${\cal C}=0$. On the contrary, ${\cal C}=1/2$ and ${\cal C}=1/3$ for the structures $2$ and $3$. On the Kagome lattice percolation of overlapping triangles takes place whereas the overlapping triangles form isolated four-site cliques on the structure 2. 

In order to investigate the relevance of the mentioned topological features first we show the prediction of the classical mean-field theory where the state is characterized by the concentration $\rho$ of cooperators. In this case the average payoff for the $C$ and $D$ strategies are
\begin{equation}
\label{eq:avpocd}
U_C= z[\rho + (1-\rho)c] \;\; \mbox{and}\;\;
U_D= z \rho b \;.
\end{equation}
The present dynamical rule, represented by the adoption probability (\ref{eq:tpo}), yields the following equation of motion for the concentration of cooperators:
\begin{eqnarray}
{\partial \rho \over \partial t} &=& \rho ( 1-\rho ) [W(D \leftarrow C) - W(C \leftarrow D)]   \nonumber \\
&=& - \rho ( 1-\rho ) \tanh \left( {U_D - U_C \over 2K } \right)
\label{eq:em2} \;.
\end{eqnarray}

According to this differential equation $\rho$ tends to zero for arbitrary value of $K$ as $U_D > U_C$. Shortly, the cooperators become extinct in those systems satisfying the conditions of mean-field approximation, e.g., if the temporal co-players are chosen randomly or in a system where all the possible pairs play a game with each other (infinite range of interaction).

Here it is worth mentioning that the cooperators also die out in the one-dimensional system \cite{szabo_pre00a} because for a confronting cooperator-defector pair the maximum cooperator's payoff ($1+c$) is always less than the minimum defector's payoff ($b$). 

For higher dimension, however, the cooperator can receive support from more than one neighboring cooperators and its total income can exceed the neighboring defector's income. For such a connectivity structure the cooperation can be sustained within a region of $b$ (and $c$) dependent on the value of noise ($K$). This work is addressed to quantify the regions of the $b$-$K$ parameter plain where cooperation can emerge. For sake of simplicity, our analysis will be restricted to the limit $c \to -0$ which is suggested by Nowak {\it et al.} in their pioneering work  \cite{nowak_ijbc93}.

\begin{figure}[ht]
\centerline{\epsfig{file=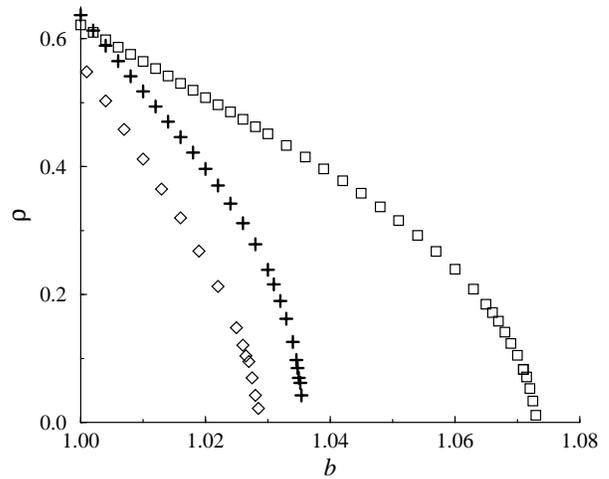,width=8cm}}
\caption{\label{fig:sqmcs}Monte Carlo results for the concentration of cooperators {\it vs}. $b$ for three different temperatures: $K=0.1$ (pluses), 0.4 (squares), and 1.2 (diamonds) on the square lattice.}
\end{figure}

 Figure \ref{fig:sqmcs} shows the concentration of cooperators on the square lattice when increasing $b$ for three different values of $K$. These data are obtained by Monte Carlo (MC) simulations performed on a block of $L \times L$ sites under periodic boundary conditions. The linear size is varied from $L=400$ to $L=2000$. The larger sizes are used in the close vicinity of the extinction of cooperators because this critical transition belongs to the so called directed percolation (DP) universality class \cite{szabo_pre98,szabo_pre00a,marro_99,hinrichsen_ap00}. 

\begin{figure}[ht]
\centerline{\epsfig{file=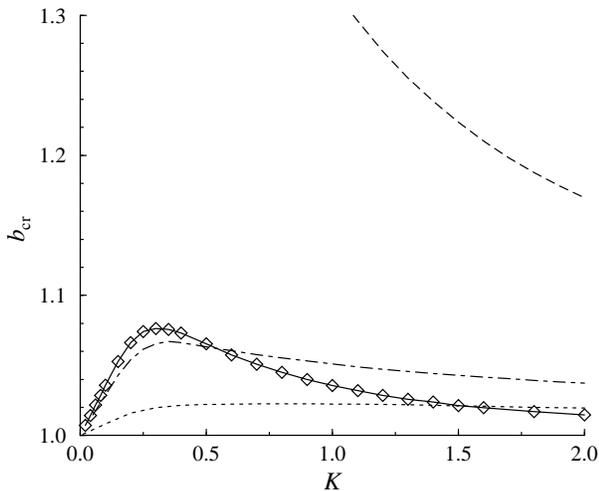,width=8cm}}
\caption{\label{fig:sqbcrt}Critical value of $b$ as a function of temperature on the square lattice. Symbols come from Monte Carlo simulations, the dashed, dotted, and dashed-dotted lines represent the prediction of generalized mean-field approximation for 2-, $2 \times 2$-, and $3 \times 3$-site clusters.}
\end{figure}

In the stationary state the concentration of cooperators is independent of the initial state and decreases monotonously if $b$ is increased. Above a threshold value ($b>b_{cr}$), however, the $C$ strategies always die out and the system remains in the homogeneous $D$ state for ever. The value of $b_{cr}$ is determined for many different values of $K$ and the results of the systematic MC simulations are summarized in Fig.~\ref{fig:sqbcrt}. Notice that $b_{cr}$ reaches its maximum value at about $K=0.32$ and $b_{cr}(K)$ tends to 1 if $K$ goes to either 0 or $\infty$. Henceforth this plot is considered as a phase diagram because the cooperators can survive only below the $b_{cr}(K)$ curve indicated by the solid line connecting the MC data in Fig.~\ref{fig:sqbcrt}.

This phase diagram differs significantly from those predicted by the above mentioned mean-field approximation [$b_{cr}^{(mf)}(K)=1$]. More adequate theoretical results are expected when using the pair approximation detailed in \cite{hauert_ajp05}. This approach is able to describe the coexistence of the $C$ and $D$ strategies, however, the value of $b_{cr}$ is significantly overestimated as shown by the dashed line (in Fig.~\ref{fig:sqbcrt}) which goes to 2 in the limit $K \to 0$. This serious shortage can be reduced by using the more sophisticated extensions of this technique when all possible configuration probabilities are determined on larger clusters. The generalization is straightforward from two-site clusters (pair approximation) to larger blocks (the essence of this method is briefly described in \cite{dickman_pra88,szabo_jpa04}). Neglecting the technical details now we report only the results of this calculation for the levels of $2 \times 2$- and $3 \times 3$-site clusters. In both cases the calculations reproduce the main qualitative features (see Fig.~\ref{fig:sqbcrt}), that is, $b_{cr}(K)$ has a maximum at a finite $K$ and $b_{cr}=1$ in the limits $K \to 0$ and $\infty$. Evidently, the accuracy of this approach is improved when choosing larger and larger clusters and the strikingly large deviations between the prediction of different levels refer to the importance of the complex short range order affected by the local topological features.  

\begin{figure}[ht]
\centerline{\epsfig{file=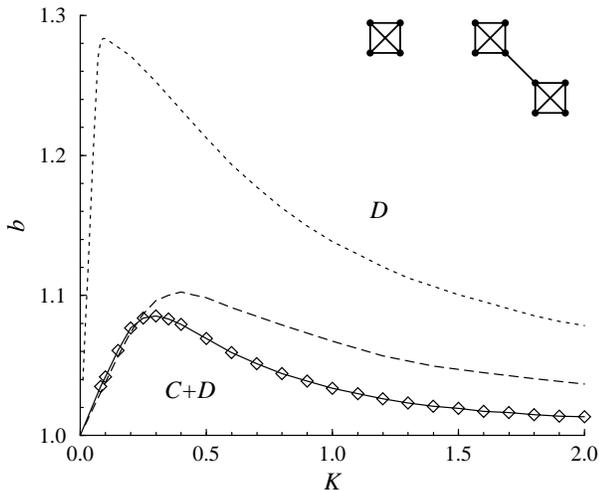,width=8cm}}
\caption{\label{fig:pd4q}$b-K$ phase diagram on the structure 2 illustrated in Fig.~\ref{fig:stages3}. Symbols denote the MC data. The dotted and dashed lines illustrate the phase boundary between the $D$ and $(C+D)$ phases as predicted by generalized mean-field approximations for the four- and eight-site clusters shown at the top.}
\end{figure}

This behavior indicates the existence of a noise level providing the highest measure of cooperation at a fixed $b$ for a square lattice connectivity structure. This means furthermore that one can observe two subsequent phase transitions (both are belonging to the DP universality class) if $K$ is increased from zero for a fixed value of temptation $b<\max (b_{cr})$. 

On structure 2 the results of MC simulations are very similar to those found on the square lattice (the differences are comparable to the symbol size) as shown in Fig.~\ref{fig:pd4q}. In contrary to the square lattice, the four-site approximation overestimates the results of MC simulations obtained on structure 2. At the same time the prediction of the eight-site approximation fits very well to the MC data for low noises ($K < 0.3$). It is suspected that the prediction of eight-site approximation (particularly for large $K$ vales) can be observed on such non-spatial structures where four-site cliques are substituted for the nodes of a random regular graph (or Bethe lattice) with a degree of four. (Notice, that the eight-site cluster is equivalent to a pair of four-site cliques and the pair approximation seems to be more correct for the Bethe lattice due to the absence of loops \cite{szabo_pre04b}).

\begin{figure}[ht]
\centerline{\epsfig{file=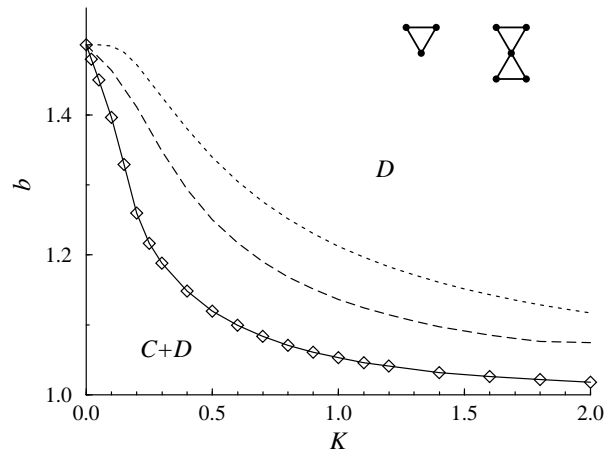,width=8cm}}
\caption{\label{fig:kagpd}Phase diagram on the Kagome lattice [structure 3 in Fig.~\ref{fig:stages3}]. Symbols denote the MC data. The dotted and dashed lines illustrate the phase boundary suggested by the three- and five-site approximation on the clusters shown at the top.}
\end{figure}

In contrary to the above phase diagrams, a qualitatively different behavior is observed on the Kagome lattice as illustrated in Fig.~\ref{fig:kagpd}. The most striking feature is that here the critical value of $b$ decreases monotonously if $K$ is increased and $b_{cr}(K=0)=3/2$ in agreement with the prediction of the three- and five-site approximations. 

In order to deduce a general picture about the relevant topological features supporting the maintenance of cooperation in the low noise limit (for the suggested dynamics) we have begun to study several other connectivity structures. According to the preliminary results the latest phase diagram (see Fig.~\ref{fig:kagpd}) is reproduced qualitatively on the square lattice with first and second neighbor interactions ($z=8$), on the triangular lattice ($z=6$), and on the body centered cubic lattice ($z=8$). In agreement with our expectation, the prediction of the five-site approximation (shown in Fig.~\ref{fig:kagpd}) is reproduced very well by the MC results obtained on the random regular structure ($z=4$) constructed from one-site overlapping triangles. For all these structures the overlapping triangles (three-site cliques) span the whole system. It would be interesting to check the emergence of cooperation (in the $K \to 0$ limit) on other networks where clique percolation takes place \cite{derenyi_prl05,palla_n05}. We have to emphasize, however, that the cooperation is not favored within the large cliques according to the mean-field arguments mentioned above. This might be another reason why the cooperation vanishes on the structure 2 in the $K \to 0$ limit. In agreement with the above conjecture the cooperators die out for vanishing $K$ on the cubic ($z=6$) and honeycomb ($z=3$) lattices. Besides it, the one-dimensional lattice with first and second neighbor interactions ($z=4$) represents an exception (because it inherits the one-dimensional features on large scales) exhibiting a sharp transition between the homogeneous states (from $C$ to $D$ if $b$ is increased for a fixed $K$).  

In summary, we have studied systematically the effect of noise $K$ (allowing irrational strategy adoptions) and temptation $b$ to choose defection on the measure of cooperation in an evolutionary Prisoner's Dilemma game for such two-dimensional lattice structures where the number of neighbors is fixed, $z=4$. For the investigated dynamical rule two basically different behaviors can be distinguished when varying the connectivity structures. In the first case the cooperators die out in the zero noise limit and the maintenance of cooperation can be optimized by choosing a suitable level of noise for any fixed value of temptation if $b<\max (b_{cr})$. In the second case the highest measure of cooperation occurs for the lowest temptation ($b=1$) and noise $K=0$ and the critical value of $b$ decreases if $K$ is increased. It is conjectured that the second behavior occurs for all the $d$-dimensional, $d \ge 2$, or non-spatial (e.g., Bethe lattice or random regular graphs) connectivity structures where the overlapping triangles span the whole system. This indicates that the percolation of the overlapping triangle in the connectivity structure can provide the optimum topological condition for the maintenance of cooperation in the situations of multi-agent Prisoner's Dilemma.

In the last years several algorithms were introduced to create a large class of networks \cite{albert_rmp02,mendes_03,newman_siamr03} and very recently the games are also suggested to control the evolution of a network \cite{zimmermann_01,biely_cm05}. The above results raise the chance that similar evolutionary PD games (in the zero noise limit) can be utilized to control the creation of networks of percolating triangles. 

\begin{acknowledgments}

This work was supported by the Hungarian National Research Fund
(Grant No. T-47003) and by the European Science Foundation (COST P10).

\end{acknowledgments}

%\bibliography{egg}

\begin{thebibliography}{35}
\expandafter\ifx\csname natexlab\endcsname\relax\def\natexlab#1{#1}\fi
\expandafter\ifx\csname bibnamefont\endcsname\relax
  \def\bibnamefont#1{#1}\fi
\expandafter\ifx\csname bibfnamefont\endcsname\relax
  \def\bibfnamefont#1{#1}\fi
\expandafter\ifx\csname citenamefont\endcsname\relax
  \def\citenamefont#1{#1}\fi
\expandafter\ifx\csname url\endcsname\relax
  \def\url#1{\texttt{#1}}\fi
\expandafter\ifx\csname urlprefix\endcsname\relax\def\urlprefix{URL }\fi
\providecommand{\bibinfo}[2]{#2}
\providecommand{\eprint}[2][]{\url{#2}}

\bibitem[{\citenamefont{Weibull}(1995)}]{weibull_95}
\bibinfo{author}{\bibfnamefont{J.~W.} \bibnamefont{Weibull}},
  \emph{\bibinfo{title}{Evolutionary Game Theory}} (\bibinfo{publisher}{MIT
  Press}, \bibinfo{address}{Cambridge, Mass.}, \bibinfo{year}{1995}).

\bibitem[{\citenamefont{Gintis}(2000)}]{gintis_00}
\bibinfo{author}{\bibfnamefont{H.}~\bibnamefont{Gintis}},
  \emph{\bibinfo{title}{Game Theory Evolving}} (\bibinfo{publisher}{Princeton
  University Press}, \bibinfo{address}{Princeton}, \bibinfo{year}{2000}).

\bibitem[{\citenamefont{Nowak et~al.}(1995)\citenamefont{Nowak, May, and
  Sigmund}}]{nowak_sa95}
\bibinfo{author}{\bibfnamefont{M.~A.} \bibnamefont{Nowak}},
  \bibinfo{author}{\bibfnamefont{R.~M.} \bibnamefont{May}}, \bibnamefont{and}
  \bibinfo{author}{\bibfnamefont{K.}~\bibnamefont{Sigmund}},
  \bibinfo{journal}{Sci. Am.} \textbf{\bibinfo{volume}{272}}
  (\bibinfo{year}{1995}).

\bibitem[{\citenamefont{Hamilton}(1964)}]{hamilton_jtb64b}
\bibinfo{author}{\bibfnamefont{W.~D.} \bibnamefont{Hamilton}},
  \bibinfo{journal}{J. Theor. Biol.} \textbf{\bibinfo{volume}{7}},
  \bibinfo{pages}{17} (\bibinfo{year}{1964}).

\bibitem[{\citenamefont{Axelrod}(1984)}]{axelrod_84}
\bibinfo{author}{\bibfnamefont{R.}~\bibnamefont{Axelrod}},
  \emph{\bibinfo{title}{The Evolution of Cooperation}}
  (\bibinfo{publisher}{Basic Books}, \bibinfo{address}{New York},
  \bibinfo{year}{1984}).

\bibitem[{\citenamefont{Hauert et~al.}(2002)\citenamefont{Hauert, De~Monte,
  Hofbauer, and Sigmund}}]{hauert_s02}
\bibinfo{author}{\bibfnamefont{C.}~\bibnamefont{Hauert}},
  \bibinfo{author}{\bibfnamefont{S.}~\bibnamefont{De~Monte}},
  \bibinfo{author}{\bibfnamefont{J.}~\bibnamefont{Hofbauer}}, \bibnamefont{and}
  \bibinfo{author}{\bibfnamefont{K.}~\bibnamefont{Sigmund}},
  \bibinfo{journal}{Science} \textbf{\bibinfo{volume}{296}},
  \bibinfo{pages}{1129} (\bibinfo{year}{2002}).

\bibitem[{\citenamefont{Nowak and May}(1993)}]{nowak_ijbc93}
\bibinfo{author}{\bibfnamefont{M.~A.} \bibnamefont{Nowak}} \bibnamefont{and}
  \bibinfo{author}{\bibfnamefont{R.~M.} \bibnamefont{May}},
  \bibinfo{journal}{Int. J. Bifurcat. Chaos} \textbf{\bibinfo{volume}{3}},
  \bibinfo{pages}{35} (\bibinfo{year}{1993}).

\bibitem[{\citenamefont{Nowak et~al.}(1994)\citenamefont{Nowak, Bonhoeffer, and
  May}}]{nowak_ijbc94}
\bibinfo{author}{\bibfnamefont{M.~A.} \bibnamefont{Nowak}},
  \bibinfo{author}{\bibfnamefont{S.}~\bibnamefont{Bonhoeffer}},
  \bibnamefont{and} \bibinfo{author}{\bibfnamefont{R.~M.} \bibnamefont{May}},
  \bibinfo{journal}{Int. J. Bifurcat. Chaos} \textbf{\bibinfo{volume}{4}},
  \bibinfo{pages}{33} (\bibinfo{year}{1994}).

\bibitem[{\citenamefont{Maynard~Smith}(1982)}]{maynard_82}
\bibinfo{author}{\bibfnamefont{J.}~\bibnamefont{Maynard~Smith}},
  \emph{\bibinfo{title}{Evolution and the theory of games}}
  (\bibinfo{publisher}{Cambridge University Press},
  \bibinfo{address}{Cambridge}, \bibinfo{year}{1982}).

\bibitem[{\citenamefont{Hofbauer and Sigmund}(1998)}]{hofbauer_98}
\bibinfo{author}{\bibfnamefont{J.}~\bibnamefont{Hofbauer}} \bibnamefont{and}
  \bibinfo{author}{\bibfnamefont{K.}~\bibnamefont{Sigmund}},
  \emph{\bibinfo{title}{Evolutionary Games and Population Dynamics}}
  (\bibinfo{publisher}{Cambridge University Press},
  \bibinfo{address}{Cambridge}, \bibinfo{year}{1998}).

\bibitem[{\citenamefont{Abramson and Kuperman}(2001)}]{abramson_pre01}
\bibinfo{author}{\bibfnamefont{G.}~\bibnamefont{Abramson}} \bibnamefont{and}
  \bibinfo{author}{\bibfnamefont{M.}~\bibnamefont{Kuperman}},
  \bibinfo{journal}{Phys. Rev. E} \textbf{\bibinfo{volume}{63}},
  \bibinfo{pages}{030901} (\bibinfo{year}{2001}).

\bibitem[{\citenamefont{Kim et~al.}(2002)\citenamefont{Kim, Trusina, Holme,
  Minnhagen, Chung, and Choi}}]{kim_pre02}
\bibinfo{author}{\bibfnamefont{B.~J.} \bibnamefont{Kim}},
  \bibinfo{author}{\bibfnamefont{A.}~\bibnamefont{Trusina}},
  \bibinfo{author}{\bibfnamefont{P.}~\bibnamefont{Holme}},
  \bibinfo{author}{\bibfnamefont{P.}~\bibnamefont{Minnhagen}},
  \bibinfo{author}{\bibfnamefont{J.~S.} \bibnamefont{Chung}}, \bibnamefont{and}
  \bibinfo{author}{\bibfnamefont{M.~Y.} \bibnamefont{Choi}},
  \bibinfo{journal}{Phys. Rev. E} \textbf{\bibinfo{volume}{66}},
  \bibinfo{pages}{021907} (\bibinfo{year}{2002}).

\bibitem[{\citenamefont{Masuda and Aihara}(2003)}]{masuda_pla03}
\bibinfo{author}{\bibfnamefont{N.}~\bibnamefont{Masuda}} \bibnamefont{and}
  \bibinfo{author}{\bibfnamefont{K.}~\bibnamefont{Aihara}},
  \bibinfo{journal}{Phys. Lett. A} \textbf{\bibinfo{volume}{313}},
  \bibinfo{pages}{55} (\bibinfo{year}{2003}).

\bibitem[{\citenamefont{Ebel and Bornholdt}(2002)}]{ebel_pre02}
\bibinfo{author}{\bibfnamefont{H.}~\bibnamefont{Ebel}} \bibnamefont{and}
  \bibinfo{author}{\bibfnamefont{S.}~\bibnamefont{Bornholdt}},
  \bibinfo{journal}{Phys. Rev. E} \textbf{\bibinfo{volume}{66}},
  \bibinfo{pages}{056118} (\bibinfo{year}{2002}).

\bibitem[{\citenamefont{Hauert and Szab{\'o}}(2005)}]{hauert_ajp05}
\bibinfo{author}{\bibfnamefont{C.}~\bibnamefont{Hauert}} \bibnamefont{and}
  \bibinfo{author}{\bibfnamefont{G.}~\bibnamefont{Szab{\'o}}},
  \bibinfo{journal}{Am. J. Phys.} \textbf{\bibinfo{volume}{73}},
  \bibinfo{pages}{405} (\bibinfo{year}{2005}).

\bibitem[{\citenamefont{Nakamaru et~al.}(1997)\citenamefont{Nakamaru, Matsuda,
  and Iwasa}}]{nakamaru_jtb97}
\bibinfo{author}{\bibfnamefont{M.}~\bibnamefont{Nakamaru}},
  \bibinfo{author}{\bibfnamefont{H.}~\bibnamefont{Matsuda}}, \bibnamefont{and}
  \bibinfo{author}{\bibfnamefont{Y.}~\bibnamefont{Iwasa}}, \bibinfo{journal}{J.
  Theor. Biol.} \textbf{\bibinfo{volume}{184}}, \bibinfo{pages}{65}
  (\bibinfo{year}{1997}).

\bibitem[{\citenamefont{Lindgren and Nordahl}(1994)}]{lindgren_pd94}
\bibinfo{author}{\bibfnamefont{K.}~\bibnamefont{Lindgren}} \bibnamefont{and}
  \bibinfo{author}{\bibfnamefont{M.~G.} \bibnamefont{Nordahl}},
  \bibinfo{journal}{Physica D} \textbf{\bibinfo{volume}{75}},
  \bibinfo{pages}{292} (\bibinfo{year}{1994}).

\bibitem[{\citenamefont{Blume}(1993)}]{blume_geb93}
\bibinfo{author}{\bibfnamefont{L.~E.} \bibnamefont{Blume}},
  \bibinfo{journal}{Games Econ. Behav.} \textbf{\bibinfo{volume}{5}},
  \bibinfo{pages}{387} (\bibinfo{year}{1993}).

\bibitem[{\citenamefont{Szab{\'o} and Hauert}(2002)}]{szabo_pre02d}
\bibinfo{author}{\bibfnamefont{G.}~\bibnamefont{Szab{\'o}}} \bibnamefont{and}
  \bibinfo{author}{\bibfnamefont{C.}~\bibnamefont{Hauert}},
  \bibinfo{journal}{Phys. Rev. E} \textbf{\bibinfo{volume}{66}},
  \bibinfo{pages}{062903} (\bibinfo{year}{2002}).

\bibitem[{\citenamefont{Hauert}(2001)}]{hauert_prsb01}
\bibinfo{author}{\bibfnamefont{C.}~\bibnamefont{Hauert}},
  \bibinfo{journal}{Proc. R. Soc. Lond. B} \textbf{\bibinfo{volume}{268}},
  \bibinfo{pages}{761} (\bibinfo{year}{2001}).

\bibitem[{\citenamefont{Schweitzer et~al.}(2003)\citenamefont{Schweitzer,
  Behera, and M{\"u}hlenbein}}]{schweitzer_acs03}
\bibinfo{author}{\bibfnamefont{F.}~\bibnamefont{Schweitzer}},
  \bibinfo{author}{\bibfnamefont{L.}~\bibnamefont{Behera}}, \bibnamefont{and}
  \bibinfo{author}{\bibfnamefont{H.}~\bibnamefont{M{\"u}hlenbein}},
  \bibinfo{journal}{Adv. Complex Systems} \textbf{\bibinfo{volume}{5}},
  \bibinfo{pages}{269} (\bibinfo{year}{2003}).

\bibitem[{\citenamefont{Szab{\'o} et~al.}(2000)\citenamefont{Szab{\'o}, Antal,
  Szab{\'o}, and Droz}}]{szabo_pre00a}
\bibinfo{author}{\bibfnamefont{G.}~\bibnamefont{Szab{\'o}}},
  \bibinfo{author}{\bibfnamefont{T.}~\bibnamefont{Antal}},
  \bibinfo{author}{\bibfnamefont{P.}~\bibnamefont{Szab{\'o}}},
  \bibnamefont{and} \bibinfo{author}{\bibfnamefont{M.}~\bibnamefont{Droz}},
  \bibinfo{journal}{Phys. Rev. E} \textbf{\bibinfo{volume}{62}},
  \bibinfo{pages}{1095} (\bibinfo{year}{2000}).

\bibitem[{\citenamefont{Szab{\'o} and T{\H{o}}ke}(1998)}]{szabo_pre98}
\bibinfo{author}{\bibfnamefont{G.}~\bibnamefont{Szab{\'o}}} \bibnamefont{and}
  \bibinfo{author}{\bibfnamefont{C.}~\bibnamefont{T{\H{o}}ke}},
  \bibinfo{journal}{Phys. Rev. E} \textbf{\bibinfo{volume}{58}},
  \bibinfo{pages}{69} (\bibinfo{year}{1998}).

\bibitem[{\citenamefont{Marro and Dickman}(1999)}]{marro_99}
\bibinfo{author}{\bibfnamefont{J.}~\bibnamefont{Marro}} \bibnamefont{and}
  \bibinfo{author}{\bibfnamefont{R.}~\bibnamefont{Dickman}},
  \emph{\bibinfo{title}{Nonequilibrium Phase Transitions in Lattice Models}}
  (\bibinfo{publisher}{Cambridge University Press},
  \bibinfo{address}{Cambridge}, \bibinfo{year}{1999}).

\bibitem[{\citenamefont{Hinrichsen}(2000)}]{hinrichsen_ap00}
\bibinfo{author}{\bibfnamefont{H.}~\bibnamefont{Hinrichsen}},
  \bibinfo{journal}{Adv. Phys.} \textbf{\bibinfo{volume}{49}},
  \bibinfo{pages}{815} (\bibinfo{year}{2000}).

\bibitem[{\citenamefont{Dickman}(1988)}]{dickman_pra88}
\bibinfo{author}{\bibfnamefont{R.}~\bibnamefont{Dickman}},
  \bibinfo{journal}{Phys. Rev. A} \textbf{\bibinfo{volume}{38}},
  \bibinfo{pages}{2588} (\bibinfo{year}{1988}).

\bibitem[{\citenamefont{Szab{\'o} et~al.}(2004)\citenamefont{Szab{\'o},
  Szolnoki, and Izs{\'a}k}}]{szabo_jpa04}
\bibinfo{author}{\bibfnamefont{G.}~\bibnamefont{Szab{\'o}}},
  \bibinfo{author}{\bibfnamefont{A.}~\bibnamefont{Szolnoki}}, \bibnamefont{and}
  \bibinfo{author}{\bibfnamefont{R.}~\bibnamefont{Izs{\'a}k}},
  \bibinfo{journal}{J. Phys. A: Math. Gen.} \textbf{\bibinfo{volume}{37}},
  \bibinfo{pages}{2599} (\bibinfo{year}{2004}).

\bibitem[{\citenamefont{Szab{\'o} and Vukov}(2004)}]{szabo_pre04b}
\bibinfo{author}{\bibfnamefont{G.}~\bibnamefont{Szab{\'o}}} \bibnamefont{and}
  \bibinfo{author}{\bibfnamefont{J.}~\bibnamefont{Vukov}},
  \bibinfo{journal}{Phys. Rev. E} \textbf{\bibinfo{volume}{69}},
  \bibinfo{pages}{036107} (\bibinfo{year}{2004}).

\bibitem[{\citenamefont{Der{\'e}nyi et~al.}(2005)\citenamefont{Der{\'e}nyi,
  Palla, and Vicsek}}]{derenyi_prl05}
\bibinfo{author}{\bibfnamefont{I.}~\bibnamefont{Der{\'e}nyi}},
  \bibinfo{author}{\bibfnamefont{G.}~\bibnamefont{Palla}}, \bibnamefont{and}
  \bibinfo{author}{\bibfnamefont{T.}~\bibnamefont{Vicsek}},
  \bibinfo{journal}{Phys. Rev. Lett.} \textbf{\bibinfo{volume}{94}},
  \bibinfo{pages}{160202} (\bibinfo{year}{2005}).

\bibitem[{\citenamefont{Palla et~al.}(2005)\citenamefont{Palla, Der{\'e}nyi,
  Farkas, and Vicsek}}]{palla_n05}
\bibinfo{author}{\bibfnamefont{G.}~\bibnamefont{Palla}},
  \bibinfo{author}{\bibfnamefont{I.}~\bibnamefont{Der{\'e}nyi}},
  \bibinfo{author}{\bibfnamefont{I.}~\bibnamefont{Farkas}}, \bibnamefont{and}
  \bibinfo{author}{\bibfnamefont{T.}~\bibnamefont{Vicsek}},
  \bibinfo{journal}{Nature} \textbf{\bibinfo{volume}{435}},
  \bibinfo{pages}{814} (\bibinfo{year}{2005}).

\bibitem[{\citenamefont{Albert and Barab{\'a}si}(2002)}]{albert_rmp02}
\bibinfo{author}{\bibfnamefont{R.}~\bibnamefont{Albert}} \bibnamefont{and}
  \bibinfo{author}{\bibfnamefont{A.-L.} \bibnamefont{Barab{\'a}si}},
  \bibinfo{journal}{Rev. Mod. Phys.} \textbf{\bibinfo{volume}{74}},
  \bibinfo{pages}{47} (\bibinfo{year}{2002}).

\bibitem[{\citenamefont{Mendes and Dorogovtsev}(2003)}]{mendes_03}
\bibinfo{author}{\bibfnamefont{J.~F.~F.} \bibnamefont{Mendes}}
  \bibnamefont{and} \bibinfo{author}{\bibfnamefont{S.~N.}
  \bibnamefont{Dorogovtsev}}, \emph{\bibinfo{title}{Evolution of Networks: From
  Biological to the Internet and WWW}} (\bibinfo{publisher}{Oxford University
  Press}, \bibinfo{address}{New York}, \bibinfo{year}{2003}).

\bibitem[{\citenamefont{Newman}(2003)}]{newman_siamr03}
\bibinfo{author}{\bibfnamefont{M.~E.~J.} \bibnamefont{Newman}},
  \bibinfo{journal}{SIAM Rev.} \textbf{\bibinfo{volume}{45}},
  \bibinfo{pages}{167} (\bibinfo{year}{2003}).

\bibitem[{\citenamefont{Zimmermann et~al.}(2001)\citenamefont{Zimmermann,
  Egu\'{\i}luz, and San~Miguel}}]{zimmermann_01}
\bibinfo{author}{\bibfnamefont{M.~G.} \bibnamefont{Zimmermann}},
  \bibinfo{author}{\bibfnamefont{V.~M.} \bibnamefont{Egu\'{\i}luz}},
  \bibnamefont{and}
  \bibinfo{author}{\bibfnamefont{M.}~\bibnamefont{San~Miguel}}, in
  \emph{\bibinfo{booktitle}{Economics and Heterogeneous Interacting Agents}},
  edited by \bibinfo{editor}{\bibfnamefont{J.~B.} \bibnamefont{Zimmermann}}
  \bibnamefont{and} \bibinfo{editor}{\bibfnamefont{A.}~\bibnamefont{Kirman}}
  (\bibinfo{publisher}{Springer Verlag}, \bibinfo{address}{Berlin},
  \bibinfo{year}{2001}), pp. \bibinfo{pages}{73--86}.

\bibitem[{\citenamefont{Biely et~al.}(2005)\citenamefont{Biely, Dragosits, and
  Thurner}}]{biely_cm05}
\bibinfo{author}{\bibfnamefont{C.}~\bibnamefont{Biely}},
  \bibinfo{author}{\bibfnamefont{K.}~\bibnamefont{Dragosits}},
  \bibnamefont{and} \bibinfo{author}{\bibfnamefont{S.}~\bibnamefont{Thurner}}
  (\bibinfo{year}{2005}), \eprint{arXiv:physics/0504190}.

\end{thebibliography}

\end{document}